\documentstyle[11pt]{article}

\begin{document}
\oddsidemargin .3in
\evensidemargin 0 true pt
\topmargin -.4in


\def\ra{{\rightarrow}}
\def\a{{\alpha}}
\def\b{{\beta}}
\def\l{{\lambda}}
\def\eps{{\epsilon}}
\def\T{{\Theta}}
\def\t{{\theta}}
\def\co{{\cal O}}
\def\car{{\cal R}}
\def\caf{{\cal F}}
\def\cs{{\Theta_S}}
\def\pr{{\partial}}
\def\tri{{\triangle}}
\def\na{{\nabla }}
\def\S{{\Sigma}}
\def\s{{\sigma}}
\def\sp{\vspace{.15in}}
\def\hs{\hspace{.25in}}

\newcommand{\be}{\begin{equation}} \newcommand{\ee}{\end{equation}}
\newcommand{\bea}{\begin{eqnarray}}\newcommand{\eea}
{\end{eqnarray}}


\begin{titlepage}
\topmargin= -.2in
\textheight 9.5in

\begin{center}
\baselineskip= 17 truept

\vspace{.3in}

\noindent
{\Large\bf Scattering of Noncommutative Strings : A Note on Signature Change
at Planck Scale}

\vspace{.4in}

\noindent
{{{\Large Supriya Kar}\footnote{skkar@physics.du.ac.in }}
{{\Large and Sumit Majumdar}\footnote{sumit@physics.du.ac.in}}}

\vspace{.2in}

\noindent
{\large Department of Physics \& Astrophysics\\
Faculty of Science, University of Delhi\\
Delhi 110 007, India}

\vspace{.2in}

{\today}
\thispagestyle{empty}

\vspace{.6in}
{\bf Abstract}

\end{center}
\vspace{.2in}

We compute the S-matrix, for the scattering of two string states, on a noncommutative $D_3$-brane
in a path integral formalism. Our analysis attempts to resolve the issue of ``imaginary string'', originally 
raised by 't Hooft in a point-particle scattering at Planck energy, by incorporating a notion of 
signature change on an emerging semi-classical D-string in the theory.

\baselineskip= 14 truept

\vspace{.3in}




\end{titlepage}

\baselineskip= 18 truept

\section{Introduction}

\subsection{Motivation}

The notion of signature change in our space-time was originally motivated by Hartle and Hawking 
\cite{hartle-hawking}, in a path integral formalism to quantum gravity. The idea was 
also independently perceived by Sakharov \cite{sakharov} at the same time. Subsequently, the novel idea involving 
a transition from Euclidean space-time to Lorentzian or vice-versa was carried forward in literatures 
over the last two decades. 

\sp

In the recent past, the issue has revived interest on a brane world-volume \cite{mars-senovilla-vera}. It is
believed that a signature change forbids the occurrence of curvature singularities in general theory of
relativity due to the fact that the Lorentzian signature is replaced by an Euclidean.
As a result, a signature change is thought of as an effective classical description, which incorporates a
quantum tunneling from Euclidean space-time to Lorentzian. Interestingly, the idea of signature change on a 
Dirichlet (D-) brane from various different perspectives has been addressed in the current literatures 
\cite{gibbons-herdeiro}-\cite{darabi}. 

\sp

In the context one of us, in collaboration \cite{kar-panda}, has
investigated a plausible signature change scenario on a 
$D_3$-brane world-volume, in its gravity decoupling limit, when the electric ({\bf E}-) field is greater than
its critical value, $i.e.$ ${\bf E}>{\bf E}_c$. It was argued that the $D_3$-brane world-volume 
changes its space-time signature from Lorentzian to 
Euclidean in the gravity decoupled theory, when ${\bf E}\rightarrow {\bf E}^+_c>{\bf E}_c$. 
This in turn sets up the notion of temperature 
at the expense of genuine time in the frame-work. It was shown that an increase in temperature leads to a series 
of successive second order phase transitions, which decouple the light gauge strings from the effective theory.
Finally the effective gauge theory on the $D_3$-brane undergoes a first order Hagedorn transition, which 
is accompanied by a further flip in signature from Euclidean to Lorentzian.  

\sp

In this article, our primary focus is on the signature of an emerging $D$-string world-sheet, in a formulation for  
a noncommutative $D_3$-brane \cite{seiberg-witten}, in the regime ${\bf E}\le {\bf E}_c$. The frame-work possesses an
underlying gravitational uncertainty principle. We develop the frame-work, systematically, 
in a open bosonic string theory in presence of an uniform two-form field in the bulk. We attempt to resolve a 
conceptual issue originally raised by 't Hooft \cite{thooft}, while considering the Planckian scattering of point 
particles in a gravitational theory. Subsequently, the issue was again put-forward by Verlinde and Verlinde \cite{ver-2} 
using scaling analysis for Planck scale physics in Einstein's theory.  
Importantly, it was shown that the two-point particle amplitude resembles to that of a string with an imaginary
string tension. It leads to an unphysical string at the Planck scale. We re-formulate the problem by
generalizing the point particle scattering at Planck energy to that of open strings. 

\sp

Our motivation behind
the mentioned generalization is multi-fold. Firstly, it is
a natural frame-work as the quantum gravitational effects are unavoidable at Planck scale and string theory 
is a candidate for a perturbative quantum theory of gravity. Though in principle, one urges for an 
ultimate nonperturabtive formulation of quantum gravity, we will see that a perturbative
string theory is good enough as a tool to explore some of the quantum gravitational nonperturbative effects. 
It is due to the fact that a string theory admits extended dynamical, nonperturbative, objects such as D-branes 
\cite{polchinski}. Secondly, the
open string frame-work uniformly treats the scattering of two string states, namely tachyons, in a ($3+1$)-dimensional
world-volume corresponding to a noncommutative $D_3$-brane. Finally, we will see that the effective noncommutative
frame-work on the $D_3$-brane naturally simplifies the scattering problem to that of forward scattering due to the
additional {\bf E}-string scale in the theory. In particular, we consider an arbitrary ${\bf E}$- and 
magnetic ({\bf B}-) fields (other than $\perp$-fields) configuration on a $D_3$-brane world-volume. 
We shall see that the frame-work incorporates two length scales, one of the order of Planck scale along the
longitudinal space and the other is a large length scale in the transverse space. Above all, our analysis 
attempts to resolve the issue of ``unphysical" string amplitude \cite{thooft,ver-2} by incorporating a plausible 
signature change on an emerging $D$-string world-sheet in an  effective noncommutative $D_3$-brane description. 
In other words, the ``unphysical'' $D$-string with an Euclidean world-sheet is correctly interpreted as a physical 
$D$-string with the Lorentzian signature. However, the Lorentzian signature of the effective space-time on the 
$D_3$-brane remains unaffected in the process. 

\sp

The plan of this article is as follows. In section 1.2, we begin with the open bosonic string dynamics and for self-sufficiency
revisit the mixed boundary conditions in theory. The set up for the two string state scattering on a $D_3$-brane is developed in 
section 2.1. and the corresponding S-matrix is obtained in 2.2. The essential theme of this article is described in section 3,
which characterizes a phase of quantum gravity using an underlying gravitational uncertainty principle in the formulation. We keep
a note on signature change on the emerging $D$-string world-sheet and attempt to resolve an issue of imaginary string observed in
a scattering phenomena for point-particles at Planck scale in section 3.3. Finally we conclude with some remarks in section 4.

\subsection{Open string modes in presence of a ${\cal B}$-field}

Consider the propagation of an Euclidean open bosonic string in $26$-dimensional Euclidean
space-time ($\mu,\nu=1,2,\dots 26$). The unitary property of S-matrix urges a closed string
propagation along with a open string in the theory. An arbitrary dynamics of a open string takes place
in presence of closed string backgrounds in its bulk, such as gravitational field 
${\cal G}_{\mu\nu}$, an antisymmetric field ${\cal B}_{\mu\nu}$ and a
scalar field. In addition, a open string is associated
with an $U(1)$ gauge field ${\cal A}_{\mu}$ at its boundary. The propagation of a open string $X^{\mu}(\tau,\s)$,
in a conformal gauge and for a constant dilaton is described by the nonlinear sigma model action 
\be
S =\ {1\over{4\pi\a'}}\int_{\S}\ d^2\s \ \left (
{\cal G}_{\mu\nu} \pr^{\tilde a} X^{\mu}\pr^{\tilde b}X^{\nu} \ 
\delta_{{\tilde a}{\tilde b}}\ -\ i 
{\cal B}_{\mu\nu}\pr_{\tilde a}X^{\mu}\pr_{\tilde b}
X^{\nu} \ \eps^{{\tilde a}{\tilde b}}\ \right )  + i \int_{\pr\S}
d\tau\ {\cal A}_{\mu}\pr_{\tau} X^{\mu} \ . \label{action-1}
\ee
Importantly, for an uniform antisymmetric closed string background, the  
${\cal B}_{\mu\nu}$ field moves to the boundary of the open string and
introduces a significant geometrical difference to the open string dynamics.
For simplicity, we restrict to a slowly varying gauge field 
${\cal A}_{\mu}(X)= - {1\over2} {\caf}_{\mu\nu}X^{\nu}$. 
A variation of the action (\ref{action-1}), for a constant metric ${\cal G}_{\mu\nu} = 
{\cal G} \delta_{\mu\nu}$, is performed to obtain the bulk equation of motion and the boundary conditions.
In a gauge choice ${\caf}_{\mu\nu}=0$, they are respectively
given by
\be
\pr^2 X^{\mu} = 0 \qquad\quad {\rm and} \quad {\cal G}_{\mu\nu} \pr_{\sigma} X^{\nu} + 
i {\cal B}_{\mu\nu}\pr_{\tau}X^{\nu} \ = 0\ , \label{action-4}
\ee
where $\partial_n$ is a normal derivative and
${\cal B}_{\mu\nu}={\cal B} \eps_{\mu\nu}$ is an invertible matrix.
The string coordinates satisfying the bulk and the boundary equations (\ref{action-4}) 
become
\be
X^{\mu}(\tau,\s) =\ X_c^{\mu}(\tau,\s)\ +\ X_m^{\mu}(\tau,\s)\ ,\label{action-5}
\ee
where $X_c^{\mu}$ and $X_m^{\mu}$, respectively, denote the conventional string coordinates and
its modification due to the $B$-field. Explicitly, with two perturbation parameters $\a'$ and 
$\big ({\cal B}^{-1}\big )^{\mu\nu}$, they are given by
\bea
&&X_c^{\mu}(\tau,\s) =\ x^{\mu} + 2\a' \ p^{\mu}\tau\ + \ i {\sqrt{2\a'}} 
\sum_{n\neq 0} {{e^{-in\tau}}\over{n}} \a^{\mu}_n\ \cos n\sigma \nonumber\\
{\rm and} &&
X_{m}^{\mu}(\tau,\s) =\ - i\big ({\cal B}^{-1}\big )^{\mu\nu}\Big ( 2\a' p_{\nu}\s 
\ +\ {\sqrt{2\a'}} 
\sum_{n\neq 0} {{e^{-in\tau}}\over{n}} \a_{{\nu}n}\
\sin n\s \Big )\ .\label{action-6}
\eea
\section{Scattering of tachyons on a $D_3$-brane}

\subsection{Path-integral formalism}

We begin with the setup to characterize an arbitrary $D_3$-brane embedded in a $26$-dimensional
space-time. A $D_3$-brane world-volume describes a ($3+1$)-dimensional space-time ($a,b=0,1,2,3$).
Since a open string end points lie on a $D_3$-brane, its dynamics can be derived in a string theory 
\cite{kar-1}. We consider a disk topology ($z = r e^{i\theta}$) for the 
open string world-sheet. The unit disk boundary
is parametrized by a polar angle ($0\le \theta \le 2\pi$). If $Y^{\mu}(y^a)$ denote the $D_3$-brane
coordinates, the required Lorentz covariant condition on the disk boundary becomes
\be
X^{\mu}(\theta)=Y^{\mu}\big ( y^a[\theta]\big )\ ,\label{tach-1}
\ee
where $y^a(\theta)$ are arbitrary functions on the disk boundary and defines the $D_3$-brane world-volume.

\sp
We adopt a path integral formalism, developed by one of us in collaboration \cite{kar-kazama}, to compute the disk amplitude
for the scattering of two tachyons on a $D_3$-brane world-volume in presence of a ${\cal B}$-field. 
Considering Dirichlet boundary conditions $i.e.$ $\pr_{\tau}X^{\mu'}=0$, for $\mu'=4,5, \dots 25$,
we reduce the $26$-dimensional tachyon interactions to that on a ($3+1$)-dimensional world-volume.
Then the problem reduces to the computation of a 2-point amplitude, for the scattering of two tachyons,
in a path-integral formalism {\cite{kar-kazama} for a $D_3$-brane. The relevant path integral for the disk amplitude
becomes
\be
{\cal V}\big ( Y^{\mu};p_1,p_2\big ) = {1\over{g_s}}\int {\cal D}X^{\mu}(z,{\bar z}) {\cal D}y^a(\theta)\
\ \delta\big ( X^{\mu}-Y^{\mu}\big )\ \exp \big (-S\big )\ \ V_1(p_1) V_2(p_2)\ ,\label{tach-2}
\ee
where $g_s$ is the string coupling constant and $S$ is the action (\ref{action-1}) with a gauge choice
${\caf}_{\mu\nu}=0$. The vertex operators for the closed string, $V_1(p_1) = 
\int d^2z_1 \exp\big ({ip_1\cdot X(z_1)}\big )$ and $V_2(p_2) = \int d^2z_2 \exp\big ({ip_2\cdot X(z_2)}\big )$, 
define the tachyon states with incoming momenta $p_1$
and $p_2$. The evaluation of path-integrals are performed uniformly over the bulk and boundary fields. The computation of
2-point amplitude, for the scattering of tachyons, on a $D_3$-brane (\ref{tach-2}) finally turns out to become
\be
{\cal V}\big ( Y^{\mu};p_1,p_2\big )= T_{D_3}\int d^3y\ dt \ {\sqrt{- \det \big (g + b \big )}}\ 
e^{ip_{\mu}Y^{\mu}}\ \int d^2z_1 d^2z_2\ \ V_{12}\ ,\label{tach-4}
\ee
where $T_{D_3}= 1/g_s (2\pi)^3\a'^2$ is the $D_3$-brane tension and $V_{12}$ is the phase factor.
In principle, the induced metric signature on a $D_3$-brane can be different from that of 
the $26$-dimensional space-time. For our purpose, we consider the metric with Lorentzian signature on the $D_3$-brane.
The induced fields $g_{ab}$ and $b_{ab}$ are 
given by $g_{ab}= {\cal G}_{\mu\nu}\pr_a X^{\mu} \pr_b X^{\nu}$ and $b_{ab}= {\cal B}_{\mu\nu}\pr_a X^{\mu} \pr_b X^{\nu}$.
The appropriate phase factor on the $D_3$-brane may be obtained from that in the bulk and is given by
\bea
V_{12}&=&\exp \Big ( {1\over2} p_{1\mu} p_{2\nu} \ {\hat e}^{\mu}_a{\hat e}^{\nu}_b\ \triangle^{ab} 
\big (z_1,z_2\big ) + 
{1\over2} p_{1\nu} p_{2\mu} \ {\hat e}^{\nu}_a{\hat e}^{\mu}_b\ \triangle^{ab} 
\big (z_2,z_1\big ) + p_{1\mu} p_{2\nu}\
{\hat e}^{\mu}_{a'} {\hat e}^{\nu}_{b'}\ D^{a'b'}\big ( z_1,z_2\big )\ 
\Big )\nonumber\\
&=&\exp \Big ({1\over2} p_{1a} p_{2b}\ \triangle^{ab}\big ( z_1,z_2\big )\ +\ 
{1\over2} p_{1b} p_{2a}\ \triangle^{ab}\big ( z_2,z_1\big )\ +\ 
p_{1a'} p_{2b'}\ \delta^{a'b'} D\big (z_1,z_2\big )\Big )\ ,\label{tach-6}
\eea
where ${\hat e}^{\mu}_a$ and ${\hat e}^{\mu}_{a'}$ are the orthonormal unit vectors, respectively, 
in the $D_3$-brane and in the transverse space to it. The
matrix propagator $\triangle^{\mu\nu}(z_1,z_2)= \left < X^{\mu}(z_1)X^{\nu}(z_2)\right >$ satisfies the bulk
equation and the boundary conditions (\ref{action-4}). They are 
\bea
&&\pr^2 \triangle^{ab}\big ( z_1,z_2\big ) = \big ( 2\pi\a'\big ) g^{ab}\ \delta^{(2)}\big 
(z_1-z_2\big ) \qquad {\rm in\; the\; bulk}\nonumber\\
{\rm and}\; && g_{ab}\ \pr_n\triangle^{ab}(z_1,z_2) - b \eps_{ab}\ \pr_t\triangle^{ab}(z_1,z_2) = 0 
\quad {\rm on\; the\; boundary,}
\label{tach-7}
\eea
where $\pr_n$ and $\pr_t$ are, respectively, the normal and tangential derivatives to the world-sheet.
The second term in the phase factor (\ref{tach-6}) is a Dirichlet function $D\big (z_1,z_2\big ) 
=\ln |z_1-z_2| -\ln |1 - z_1{\bar z}_2|$.
The (inverse) matrix propagator satisfying the eqs.(\ref{tach-7}) for its diagonal elements is given by
\be
\triangle^{ab}\big ( z_1,z_2\big )|_{a=b} = \a'g^{ab}\ln |z_1-z_2| + \ {\a'} 
\left ( {1\over{g-b}}\cdot \big ( g+bg^{-1}b\big )\cdot {1\over{g+b}}\right )^{ab}\ 
\ln \left |1-{1\over{z_1{\bar z}_2}}\right |
\ .\label{tach-10}
\ee
The off-diagonal elements are worked out to yield
\be
\triangle^{ab}\big ( z_1,z_2\big )|_{a\neq b} = {\a'} 
\left ( {1\over{g-b}}\cdot b\cdot {1\over{g+b}}\right )^{ab}\ 
\ln \left ({{z_1-{\bar z}_2}\over{{\bar z}_1- z_2}} \right )
\ .\label{tach-11}
\ee
\subsection{S-matrix}

Now consider a conformal map from the disk topology to an upper-half of the complex plane.
Then the disk boundary is parametrized by its real value $\tau$. On the boundary, 
$i.e.$ on a $D_3$-brane, the diagonal (\ref{tach-10}) and off-diagonal (\ref{tach-11}) elements are simplified 
to yield a generalized matrix propagator. It is given by
\be
\triangle^{ab}\big ( y,y'\big )= \left < Y^a(y) Y^b(y')\right >
= 2\a'G^{ab}\ \ln (y-y')\ +\ {i\over2} \Theta^{ab}\ \eps \big ( y -y'\big ) \ ,\label{tach-12}
\ee
where $\eps(y-y')$ is step function, $G^{ab}$ and $\Theta^{ab}$ are respectively the modified metric and a simplectic two-form 
on the $D_3$-brane. They are given by
\be
G^{ab} = \left ( {1\over{g-b}}\cdot g\cdot {1\over{g+b}}\right )^{ab} \qquad
{\rm and}\quad \Theta^{ab}= (2\pi\alpha') \left ( {1\over{g-b}} \cdot b \cdot {1\over{g+b}}\right )^{ab}\ .\label{tach-13}
\ee
Since $D(\tau,\tau')=0$, the scattering matrix (\ref{tach-6}) on the $D_3$-brane becomes
\bea
V_{12}(p_{1},p_{2};y^1,y^2)&=&\exp \Big ( 2\a'\ p_{1a} p_{2b}\ G^{ab} \ln \big (y^1-y^2\big ) + {i\over2} p_{1a}p_{2b}
\ \Theta^{ab} \ \eps\big (y^1-y^2\big )\Big )\nonumber\\
&=&\big (y^1-y^2\big )^{2\a'p_1\cdot p_2}\ \exp \Big ({i\over2} p_{1a}p_{2b}
\ \Theta^{ab} \ \eps\big (y^1-y^2\big )\Big )
\ .\label{tach-14}
\eea
The effective space-time geometry on the $D_3$-brane world-volume may be derived using the matrix
propagator in eq.(\ref{tach-12}). It is a simple check to note that the space-time coordinates 
are noncommutative, $i.e.$ $\big [ Y^a(y) , Y^b(y')\big ] = i \Theta^{ab} \eps ( y-y')$.

\section{Nonperturbative effects}

\subsection{Effective dynamics on a $D_3$-brane}

The action describing the dynamics of a $D_3$-brane may be derived from the
scattering amplitude (\ref{tach-4}) for $p_1=0=p_2$. It is straight-forward to see that the $D_3$-brane 
dynamics is governed by the Born-Infled (BI-) action and with a Lorentzian signature the action 
becomes 
\be
S_{D_3}= T_{D_3}\int dt\ d^3y \ {\sqrt{-\Big (g + b + {\bar F}\Big )}}\ ,\label{np-1}
\ee
where ${\bar F}_{ab}=(2\pi\a') F_{ab}$. In addition to an induced metric $g_{ab}$, the $D_3$-brane world-volume
admits an $U(1)$ gauge theory. The electromagnetic (EM-)
field strength in ($3+1$)-dimensions is expressed in terms of its electric $E_s={\caf}_{0s}$ 
and magnetic $B_{s_1} =\eps_{s_1s_2s_3} {\caf}_{s_2s_3}$ components. 
Since the direction of {\bf E}-field is all along the open string,
an electric stretch is either parallel or anti-parallel to that of its orientation. 
The stretch builds up a string tension. It introduces the notion of an ${\bf E}$- or 
EM-string \cite{kar-panda}, which is independent of the conventional open string in the theory. 
This phenomenon is unlike to the one due to the magnetic field, since a {\bf B}-field rather
introduces a width transverse to the direction of propagation. So a magnetic field does not 
contribute to the string tension. Thus the effective string tension for the noncommutative 
$D_3$-brane becomes ${T}^2_{\rm eff}=T^2_{D_3}- {\bf E}^2$.

\sp

In this article, we focus on the effective noncommutative space-time
description on the $D_3$-brane. Seiberg-Witten map \cite{seiberg-witten} is used to write the BI-action 
(\ref{np-1}) in terms of an $U(1)$ noncommutative gauge field for a fixed $\Theta$. 
Under the map, the closed string
parameters are transformed to that of the noncommutative open string, $i.e.$ $g_s\rightarrow G_s$ and 
$g_{ab}\rightarrow G_{ab}$. In addition the $b_{ab}$ is used to define the noncommutative geometry  
with $\Theta^{ab}$ parameter in the effective $D_3$-brane dynamics. As a result, the gauge field
strength is transformed to that of a noncommutative, $i.e.$ $F_{ab}\rightarrow {\hat F}_{ab}$. 
Then the $D_3$-brane dynamics (\ref{np-1}) is governed by the corresponding noncommutative BI-action. 
It is given by 
\be
S_{D_3}= {\hat T}_{D_3}\int d^3y\ dt \ {\sqrt{-\Big (G + (2\pi\a') {\hat F}\Big )}}\ ,\label{np-51}
\ee
where ${\hat T}_{D_3}= (g_s/G_s)T_{D_3}$. The $U(1)$ noncommutative field strength is expressed as
${\hat F}_{ab}=\pr_a{\hat A}_b - \pr_b{\hat A}_a - i\big [{\hat A}_a , {\hat A}_b\big ]$. At this point,
it may be interesting to analyze the scattering phenomenon for the gauge particles at Planck scale 
\cite{kar-maharana} on a noncommutative $D_3$-brane. However, the detail analysis is beyond the scope of 
this article.

\subsection{Two different length scales}

Now let us consider an arbitrary field configuration ${\bf E}= 
(0,E_2,E_3)$ and ${\bf B}=(0,B_2,B_3)$. For simplicity, we take 
$E_2=E_3=E$ and $B_2= B_3=B$.  The mode expansion for the open string
coordinates (\ref{action-6}) ending on a noncommutative $D_3$-brane world-volume is worked out.
Using the Lorentz covariant condition (\ref{tach-1}), the $D_3$-brane coordinates are given by
\bea
&&Y^0\big (y^a[\tau]\big ) = \ y^0  + 2\a'\ p_0 \tau + i{\sqrt{2\a'}} \sum_{n\neq 0}\ 
{{e^{-in\tau}}\over{n}} (-1)^n\ \a_{0n}  \ +\ \theta_E \big ( p_2 + p_3\big ) \ ,
\nonumber\\
&&Y^1\big (y^a[\tau]\big ) = \ y^1  + 2\a'\ p_1 \tau + i{\sqrt{2\a'}} \sum_{n\neq 0}\ 
{{e^{-in\tau}}\over{n}} (-1)^n\ \a_{1n}  \ -\ \theta_B\big ( p_2 - p_3\big ) \ ,
\nonumber\\
&&Y^2\big (y^a[\tau]\big ) = \ y^2  + 2\a'\ p_2 \tau + i{\sqrt{2\a'}} \sum_{n\neq 0}\ 
{{e^{-in\tau}}\over{n}} (-1)^n\ \a_{2n}  \ +\ \big (\theta_E p_0 + \theta_B p_1\big ) 
\nonumber\\
{\rm and}\quad 
&&Y^3\big (y^a[\tau]\big ) = \ y^3  + 2\a'\ p_3 \tau + i{\sqrt{2\a'}} \sum_{n\neq 0}\ 
{{e^{-in\tau}}\over{n}} (-1)^n\ \a_{3n}  \ +\ \big (\theta_E p_0 - \theta_B p_1\big ) \ ,
\label{np-8}
\eea
where $\theta_E= - E^{-1}$ and $\theta_B= -B^{-1}$ are, respectively, the electric and 
magnetic noncommutative parameters. They may be 
obtained in the limit $g\rightarrow 0$ from $\Theta^{ab}$ in eq.(\ref{tach-13}).
Using notations in eq.(\ref{action-5}), these modes may be re-expressed as
\bea
&&Y^0 =\ Y_c^0 \ -\ \theta_E \big ( p_2+p_3\big )  \ ,\nonumber\\
&&Y^1 =\ Y_c^1 \ -\ \theta_{\cal B}\big (p_2-p_3\big ) \ ,\nonumber\\
&&Y^2 =\ Y_c^2 \ +\ \big ( \theta_E p_0 + \theta_B p_1\big )\nonumber\\ 
{\rm and}\quad 
&&Y^3 = \ Y_c^3 \ +\ \big (\theta_E p_0 - \theta_B p_1\big ) \ .
\label{np-9}
\eea
It implies that the $D_3$-brane coordinates receive corrections only from the constant momentum modes 
associated with the noncommutative parameters $\theta_E$ and $\theta_B$ in the theory.  In other words,
the trajectory of a open string on a noncommutative $D_3$-brane receives a constant shift. The hamiltonian may be seen to receive corrections due to the additional momentum modes in the theory. It may be given by
\be
L^b_0=\a'p^2 - \a' \big ( bg^{-1}b\big )_{ab} p^ap^b + N_{\rm osc}
\ ,\label{np-10}
\ee
where $N_{\rm osc}$ is the number operator with its eigen values ($0,1,2,\dots $) in the theory. 
The mass-shell condition $\big (L^b_0-1\big )\left |
{\rm phy}\right >=0$ is worked out for a single winding state to 
show that the radius of time-like coordinate $Y^0$ is of
the order of string or Planck scale $R={\sqrt{\a'}}=l_s$ \cite{kar-panda}. 
Though the result is natural in the frame-work, it helps us to comment on the nature of other 
$3$-space like coordinates on the$D_3$-brane using the inherent noncommutativity in the
theory. In fact, one may argue to show that the radii of the transverse 
coordinates, $Y^2$ and $Y^3$, specified by a length scale $l_{\perp}$ are large $i.e.$ $l_{\perp}>>l_s$. 
For instance, a non-vanishing $\Theta_E$ in the commutation relations 
confirms the above assertion. Taking into account the presence of $\Theta_{\cal B}$, assures that 
the radius for $Y^1$ is also small (order of $l_s$). The presence of two distinct scales ($l_s>>l_{\perp}$) 
on a noncommutative $D_3$-brane is natural in our frame-work. In fact, the noncommutativity, among the space-time coordinates, 
apparently incorporates a gravitational uncertainty principle. In the context, t'Hooft has proposed the construction of
S-matrix for the black hole formation and evaporation using the above underlying principle \cite{thooft}. A schematic 
view of space-time on a $D_3$-brane world-volume is depicted in fig.\ref{kmfig1}. 

\begin{figure}[ht]
\begin{center}
\vspace*{3in}
\relax{\includegraphics{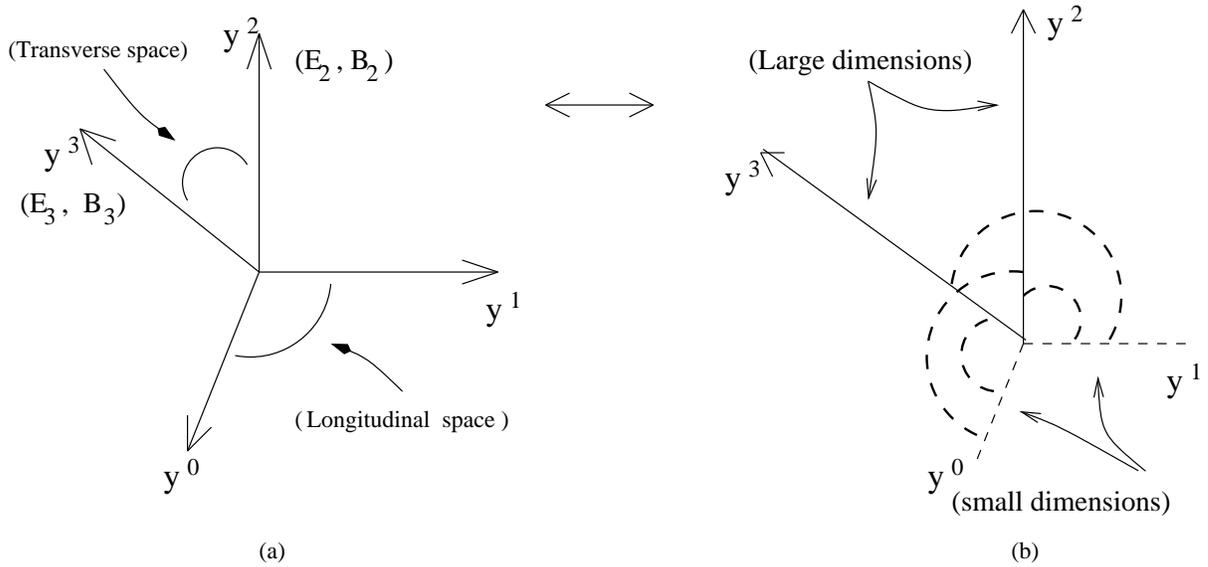}}
\end{center}
\vspace*{-.1in}
\caption{Space-time on a $D_3$-brane in presence of an EM-field. Fig.(a) illustrates an ordinary or commutative
space-time. Fig.(b) a noncommutative space-time obtained by a
Seiberg-Witten map from (a). Different length scales $i.e.$ small scales on the longitudinal 
plane ($y^0$ and $y^1$) and large scales on the transverse space ($y^2$ and $y^3$) are schematically shown in fig.(b).}
\label{kmfig1}
\end{figure}

\sp

Now let us characterize the ($3+1$)-dimensional space-time on the $D_3$-brane in terms of longitudinal space 
($\a,\beta=0,1$) and transverse space ($i,j=2,3$). The small length scale on the longitudinal space introduces a
periodicity among its coordinates, $i.e.$ $Y^0\rightarrow Y^0+ m(2\pi R_0)$ and $Y^1\rightarrow Y^1 + n(2\pi R_1)$, 
for integer values of $m,n$. It generates a time-like killing vector ($\pr/\pr Y^0$) in addition to 
a space-like ($\pr/\pr Y^1$) one. In other words, the continuity along the longitudinal coordinates $Y^{\a}$ are 
approximated by a large number ($m,n\rightarrow \infty$) of tiny intervals ($2\pi R_0$) and ($2\pi R_1$). It
leads to translation symmetries along $y^0$ and $y^1$ on the $D_3$-brane world-volume, $i.e.$ 
$Y^a\big (y^{\a}y^i\big )\rightarrow Y^a\big (y^2,y^3\big )$. The periodic time like coordinate implies that
the very notion of time looses its identity \cite{atick-witten}. Naively, a signature flip from Lorentzian to 
Euclidean is advocated in the theory, without any loss of dimension of space-time. However the Euclidean signature
is achieved in the usual way, $i.e.$ by Wick rotating the time coordinate. We will see that the signature change is
enforced on the emerging $D$-string world-sheet by the consistency of the theory, though the signature on the noncommutative
$D_3$-brane world-volume remains unaffected. 

\sp
\noindent
Then the matrix propagator (\ref{tach-12}) in the effective description becomes
\be
\triangle^{ab}\big ( y^2,y^3\big ) = 2\a' G^{ab} \ln \big (y^2 -y^3\big ) \ +\ 
{i\over2} \Theta^{ab} \eps \big (y^2-y^3\big )\ .\label{np-11}
\ee
It means that the 
noncommutative $D_3$-brane wraps the small dimensions $Y^0$ and $Y^1$ around the transverse space ($Y^2,Y^3$) and
leaves behind a semi-classical $D$-string, with a large winding number. Since $\Theta^{23}=0$, the emerging $D$-string in the
transverse space possesses an ordinary geometry with an induced metric $h_{ij}$. Interestingly, the $D$-string in the context
is identical to that obtained for the point-particle scattering at Planck energy \cite{thooft,ver-2}. It corresponds to a 
gauge choice $G_{\a i}=0$ for the ($3+1$)-dimensional effective metric
\be
G_{ab}= \pmatrix {{{\bar h}_{\a\beta}} & {0} \cr {0} & {h_{ij}}}\ ,\label{np-12}
\ee
where ${\bar h}_{\a\beta}$ is the metric on the longitudinal space. 
In addition to $h_{ij}$, the $D$-string describes the dynamics of an U(1) gauge field $A_i$. 

\subsection{Signature change on the emerging $D$-string}

Now the propagator (\ref{np-11}) in the transverse space on the $D$-string becomes 
\be
\triangle^{ij}\big ( y^2,y^3\big ) = 2\a' \ h^{ij} \ \ln \big (y^2 -y^3\big ) \ .\label{np-13}
\ee
The S-matrix (\ref{tach-14}) for the two-tachyon scattering on a noncommutative $D_3$-brane takes a final
form
\be
V_{12}\big ( p_1,p_2;y^2,y^3\big )=\exp \Big (p_{1i}p_{2j}\ \triangle^{ij}\big ( y^2,y^3\big )\Big )\ .\label{np-14}
\ee
At this point we re-emphasize that our frame-work is a natural one at Planck scale and is a stringy generalization 
of two-particle scattering amplitude obtained by Verlinde and Verlinde \cite{ver-2}. 
The small length scale $l_s$ along $Y^0$ and $Y^1$ naturally invokes a forward scattering of two tachyons in a
($3+1$)-dimensional noncommutative space time on a $D_3$-brane world-volume. Since the mass of the interacting 
particles is very small in comparison to that of the Planck scale,  they may be treated as 
left (L) and right (R) moving tachyons in the theory. Then the trajectory for the tachyons or string states
(\ref{np-9}) may be given by
\bea
&&Y^{L}_{\rm in} =\ Y_c^L \ -\ p_L^{\perp}\ \theta^L\ \eps \big (y-y_L\big )  \nonumber\\
{\rm and} &&Y^R_{\rm in} =\ Y_c^R \ -\ p_R^{\perp}\ \theta^R\ \eps \big (y-y_R\big )\ ,
\label{np-14}
\eea
where $p_L^{\perp}=(p_2-p_3)_L$ and $p_R^{\perp}=(p_2-p_3)_R$ are, respectively, the small transverse momenta 
of $L$- and $R$- moving tachyons. $\theta^L= \big (\theta_E +\theta_B\big )$ and $\theta^R= \big (\theta_E -\theta_B\big )$ 
are the effective noncommutative parameters in the theory. The trajectories (\ref{np-14}) for both the tachyons contain 
a discontinuity. They may be interpreted as a generalization of the shock wave solutions in 
Einstein equations obtained in \cite{ver-2}. A schematic view of the scattering of two string states in the frame-work 
is shown in fig.\ref{kmfig2}.  

\begin{figure}[ht]
\begin{center}
\vspace*{2.6in}
\relax{\includegraphics{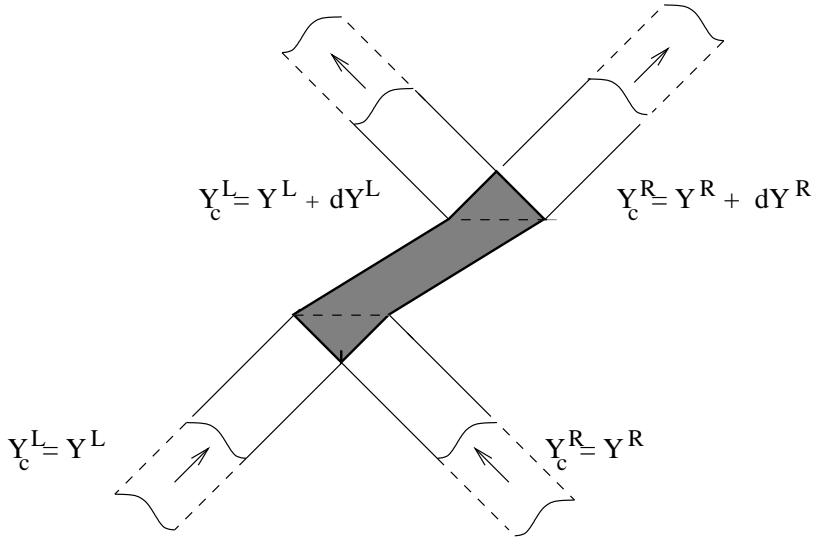}}
\end{center}
\vspace*{-.1in}
\caption{Nontrivial shifts ($dY^L$ and $dY^R$) 
in the trajectories for the scattering of $L$- and $R$- moving strings in an effective ($3+1$)-dimensional 
noncommutative space-time on a $D_3$-brane. It describes a simultaneous shock wave geometry
due to the discontinuities along its stringy trajectory.}
\label{kmfig2}
\end{figure}

The shift in geodesics for both the $L$- and $R$- moving conventional strings is a new phenomenon in the frame-work. 
It shows that the world-sheet for both the strings at the interaction vertex
receives an instantaneous shift (\ref{np-14}) simultaneously along the internal directions, $i.e.$ transverse to the 
longitudinal space. The incoming string momenta are worked out to yield
\bea  
&&P^L_{\rm in} = p^L\ \delta \big (y-y^L\big )\nonumber\\
{\rm and} &&P^R_{\rm in} = p^R\ \delta \big (y-y^R\big )\ .\label{np14}
\eea
Though the phenomenon is a generalization to that of the point-particle, the resulting 
simultaneous shock wave geometry is new in the frame-work.  A  semi-classical $D$-string is argued to propagate 
in the transverse space on an effective noncommutative $D_3$-brane (\ref{np-51}) with the metric signature ($ - + + + $). 
The small and large length scales in the theory lead to a quantum gravity 
phase in the longitudinal space ($y^0,y^1$) and a semi-classical gravity dynamics in the transverse space. Since the
genuine time looses its identity in presence of an {\bf E}-field, the metric signature may also be viewed either as 
($+ + + -$) or ($+ + - +$) without any change in signature on the $D_3$-brane. The apparent degeneracy in the metric
signature is consistent with the quantum gravity phase in the effective theory. In other words, the string scale radius
for the a priori time-like coordinate $Y^0$ may be viewed intuitively as internal flips: $Y^0\rightarrow Y^4$ and $Y^3$ 
(or $Y^2)\rightarrow {\tilde Y}^0$ a time-like coordinate. Then the action (\ref{np-51}) in the effective description 
reduces to that for a $D$-string with windings. It takes the form
\bea
S_{D_3}&=&T_{D_3}\int dy^4 dy^1 \int dy^2 d{\tilde y}^0\ {\sqrt{\det\Big ({\tilde h} + {\bar F}\Big )_{\a\beta}}}
\ {\sqrt{- \det \Big (h+{\bar F}\Big )_{ij}}}
\nonumber\\
&\equiv & T_{D_1}\int dt\ dy^2 \ {\sqrt{-\det \Big (h+{\bar F}\Big )_{ij}}}\ .\label{np-141}
\eea
The $D$-string world-sheet possesses a Lorentzian metric signature ($- +$) in the frame-work. As can be seen from eq.(\ref{np-141}),
a simple physical tool leads to a change in signature from the Euclidean to Lorentzian on the $D$-string. The intuitive interpretation 
possibly resolves the issue of imaginary string tension discussed in the literatures \cite{thooft,ver-2} while investigating 
the point-particle scattering phenomenon at Planck scale. It was pointed out that a 2-point scattering 
amplitude in Einstein's theory when raised to Planck scale apparently resembles to that obtained in a string theory with 
an unusual factor of $i$ in front of the string action with an Euclidean signature. 

\sp

Naively, the signature change appears as a trick. However, we will see that it requires an intuitive 
understanding of the complete scenario. In fact, it is due to the ${\bf E}$-string, which 
is associated with some of the unusual features unlike to that of a conventional string.
It means that the time coordinate on a noncommutative $D_3$-brane mixes with the space coordinates \cite{kar-panda} and 
one needs to re-define all four coordinates. The signature change is provocative to understand the small radius
for the time-like coordinate on the noncommutative $D_3$-brane.

\section{Concluding remarks}

In this article, we have systematically investigated the scattering of two strings on a noncommutative $D_3$-brane. In
particular, we have considered the scattering of two tachyons in a open bosonic string theory with appropriate Dirichlet
boundary conditions. Our frame-work provides a natural forum to study a forward scattering phenomenon in quantum 
gravity. The noncommutative $D_3$-brane coordinates in the theory set up a gravitational uncertainty principle, which is
shown to incorporate two independent length scales in the theory. The emergence of a small length scale dictate
a nonperturbative quantum gravity phase in the theory which is governed by a strong coupling constant. The large length
scale in the theory elopes in the transverse space and is shown to describe a semi-classical $D$-string. It is argued that
the $D$-string world-sheet undergoes a change in signature from Euclidean to Lorentzian without changing the overall signature 
of ($3+1$)-dimensional space-time. The degenerate metric signature on the transverse space is enforced by the quantum gravity
phase in the theory.

\sp

Our frame-work was primarily motivated to resolve an issue of ``imaginary string'' 
originally raised by t'Hooft \cite{thooft} and subsequently by Verlinde and Verlinde \cite{ver-2} while investigating the
Planckian scattering of point-particles in Einstein's theory. The string set up provides a general and a natural frame to investigate
the forward scattering in quantum gravity. Our result for the scattering amplitude is in agreement with that obtained for the point
particles. In the process, we noticed that the trajectories of two string states 
receive simultaneous constant shifts, which is not possible in a point-particle scattering phenomenon. The reason being that 
the shifts causing gravitational shock waves in the frame-work are transverse to the propagation of incoming strings. Our analysis
is an attempt to resolve the issue of ``imaginary string'' by conjecturing a change in signature on the emerging $D$-string in a 
noncommutative $D_3$-brane. Since the set up is described with an inherent gravitational uncertainty principle, it may allow a
forum to construct an appropriate S-matrix by taking into account the succesive black hole creation and evaporation.

\sp
\sp

\noindent
{\large\bf Acknowledgments}

\sp

We acknowledge the discussions with the participants in the International workshop on Strings, 
15-23 December 2004, at Khajuraho, India, where an initial version of this work was presented. In particular, 
we are grateful to Gary W. Gibbons, Bala Sathiapalan, Ashoke Sen and Erik Verlinde for their helpful comments.
S.K. acknowledges a partial support, under SERC fast track young scientist PSA-09/2002, from the
D.S.T, Govt.of India. The work of S.M. is partly supported by a C.S.I.R. research fellowship.

\vfil\eject

\def\anp{Ann. of Phys.}
\def\prl{Phys. Rev. Lett.}
\def\prd#1{{Phys. Rev.} {\bf D#1}}
\def\jhep{J.High Energy Phys.}
\def\cqg#1{{Class. \& Quantum Grav.}}
\def\plb#1{{Phys. Lett.} {\bf B#1}}
\def\npb#1{{Nucl. Phys.} {\bf B#1}}
\def\mpl#1{{Mod. Phys. Lett} {\bf A#1}}
\def\ijmpa#1{{Int. J. Mod. Phys.} {\bf A#1}}
\def\rmp#1{{Rev. Mod. Phys.} {\bf 68#1}}


\end{document}